\documentclass[prl,twocolumn,showpacs]{revtex4}
\usepackage{graphicx}

\begin{document}

\title{Efficient Statistical Inference for Stochastic Reaction
  Processes}
\author{Andreas Ruttor}
\author{Manfred Opper}
\affiliation{Artificial Intelligence Group, TU Berlin, Berlin,
  Germany}
\date{June 29, 2009}

\begin{abstract}
  We address the problem of estimating unknown model parameters and
  state variables in stochastic reaction processes when only sparse
  and noisy measurements are available. Using an asymptotic system
  size expansion for the backward equation, we derive an efficient
  approximation for this problem. We demonstrate the validity of our
  approach on model systems and generalize our method to the case when
  some state variables are not observed.
\end{abstract}

\maketitle

Stochastic reaction processes are models for the dynamics of a
population of interacting species, e.g., chemical substances which can
participate in several reactions. They play an important role in
physics, chemistry, and biology \cite{vanKampen:1981:SPP,
  Gardiner:1996:HSM}. Given the reaction rates of the process, the
probability of having a specific number of individuals (molecules) in
the system at a certain time is computed by solving the chemical
Master equation for which efficient exact and approximate approaches
are known.

In recent years, this modelling framework has found increasing
interest in the field of systems biology, where it has been applied to
gene regulation and protein networks \cite{Wilkinson:2006:SMS}. The
fact that here reactions are known only qualitatively leaving a
variety of numerical rate parameters unknown leads to a different type
of computational problem. It becomes necessary to estimate these
parameters using discrete noisy observations of the stochastic
process. The information contained in the observations also changes
the time evolution of the system's probability distribution. This
problem of state and parameter estimation is known as data
assimilation \cite{Ide:2007:DA}.

When observations are frequent and the process is in equilibrium, one
may estimate parameters directly by fitting the stationary
distribution or correlation functions of the process to their
counterparts measured from the data. For an application of this method
to models of financial markets, see \cite{Alfarano:2004:EAB}. But the
problem of statistical inference becomes especially nontrivial when
observations are sparse. In this case, one would like to apply
statistically efficient methods such as maximum-likelihood (ML)
estimators or Bayes estimators. Unfortunately, exact computations of
such estimators or their simulation by stochastic techniques such as
particle methods or Monte Carlo approaches may become time consuming
or intractable \cite{Golightly:2005:BIS, Doucet:2003:PEG,
  Boys:2008:BID}. To cope with this problem, we present in this Letter
an efficient approximate solution based on an asymptotic system size
expansion \cite{vanKampen:1981:SPP}.

We begin by formulating a stochastic reaction process for $d$
interacting species, which participate in several chemical reactions.
The statistical dynamics for the state $\mathbf{x} = (x_1, \ldots,
x_d)$ where $x_j$ denotes the numbers of individuals (molecules) of
species $j$, is governed by a continuous time Markov jump process. The
rate of each reaction $i$ is given by its rate function
$h_i(\mathbf{x})$. Waiting times between two consecutive reaction
events are exponentially distributed with rate parameter
$\lambda(\mathbf{x}) = \sum_i h_i(\mathbf{x})$. At each time such an
event occurs, a specific reaction $i$ is chosen with probability
$h_i(\mathbf{x}) / \lambda(\mathbf{x})$ and the state $\mathbf{x}$
changes deterministically into some other state $\mathbf{x}'$ which
depends on the reaction.

The probability $p_t(\mathbf{x})$ of finding the system in state
$\mathbf{x}$ at any time $t$ evolves through the Master equation
\begin{equation}
  \frac{d}{dt} p_t(\mathbf{x}) = \sum_{\mathbf{x}' \neq \mathbf{x}}
  \left[ p_t(\mathbf{x}') f_\theta(\mathbf{x} | \mathbf{x}') -
    p_t(\mathbf{x}) f_\theta(\mathbf{x}' | \mathbf{x}) \right] \,,
  \label{eq:master}
\end{equation}
where the total rate $f_\theta\left(\mathbf{x}'|\mathbf{x}\right)$ is
the sum of the rates $h_i(\mathbf{x})$ for all processes which lead
from $\mathbf{x}$ to $\mathbf{x}'$. $\theta$ denotes a vector of
parameters which determine the explicit form of the $h_i$. This might
be, e.g., the rate constants appearing in the mass action kinetics of
chemical reactions \cite{Guldberg:1879:CCA}.

We will now address the problem of estimating the parameters $\theta$
from a set of measurements $D \equiv \{\mathbf{y}_i\}_{i=1}^N$. The
$\mathbf{y}_i$ are noisy observations of the true process
$\mathbf{x}(t_i)$ at discrete times $t_i$. A related problem is the
computation of $p_t(\mathbf{x}|D)$, the probability of the state at
time $t$ based on past and future observations.

Statistically efficient methods for parameter estimation can be based
on the likelihood of the observations $p(D|\theta)$. A maximum
likelihood (ML) approach would maximize $p(D|\theta)$ with respect to
$\theta$, whereas a Bayesian procedure would compute a posterior
density of parameters $p(\theta|D) \propto p(D |\theta) \, p(\theta)$
when prior knowledge in the form of a density $p(\theta)$ is
available.

It is straightforward to derive a recursion for $p(D|\theta)$. For any
time $t$, we define $r_t(\mathbf{x}) \equiv p(D_{\geq t} | \theta,
\mathbf{x}_t = \mathbf{x})$, the likelihood of future observations
$D_{\geq t} = \{\mathbf{y}_i\}_{t_i \geq t}$ conditioned on the
present state $\mathbf{x}(t) = \mathbf{x}$. The likelihood of all data
is then $p(D |\theta) = \sum_{\mathbf{x}} p_0(\mathbf{x}) r_0
(\mathbf{x})$, where $p_0(\mathbf{x})$ is the distribution of the
initial state. For times between two observations, $r_t$ obeys the
Kolmogorov backward equation
\begin{equation}
  \frac{d}{dt} r_t(\mathbf{x}) = \sum_{\mathbf{x}'\neq \mathbf{x}}
  f(\mathbf{x}' | \mathbf{x}) \left[ r_t(\mathbf{x}) -
    r_t(\mathbf{x}') \right] \,.
  \label{eq:backward}
\end{equation}
Here and in the following, we omit the parameters $\theta$ in the rate
$f$ for notational clarity. The observations enter through their
conditional distributions (assuming independent noise)
$p(\mathbf{y}|\mathbf{x})$, in the end condition $r_{t_N}(\mathbf{x})
= p(\mathbf{y}_N | \mathbf{x}(t_N))$ and in the jump conditions
\begin{equation}
  \lim_{t \rightarrow t_l^-} r(\mathbf{x}, t) = p(\mathbf{y}_l |
  \mathbf{x}(t_l)) \lim_{t\rightarrow t_l^+} r(\mathbf{x}, t) \,,
  \label{eq:jumpcond}
\end{equation}
where $t_l^-$ and $t_l^+$ denote the left and right side limits. A
numerical inference method would process a sequence of parameter
values $\theta$ together with their likelihoods $p(D |\theta)$ either
for optimization or for drawing samples from the posterior. For each
of these values, a full solution of Eq.~(\ref{eq:backward}) backwards
in time is required.

Given the exact parameters or good estimates, the conditional
probability $p_t(\mathbf{x} | D)$ is obtained using the Markov nature
of $\mathbf{x}(t)$ and Bayes' rule as $p_t(\mathbf{x} | D) \propto
p_t(\mathbf{x} | D_{< t}) \, r_t(\mathbf{x})$, where $p_t(\mathbf{x} |
D_{< t})$ is the conditional distribution of the state based only on
the observations $D_{< t} \equiv \{\mathbf{y}_i\}_{t_i < t}$ before
time $t$. For times between observations, this probability fulfils the
forward Eq.~(\ref{eq:master}) with jump conditions at the
observations. This result can be further simplified by noting that the
conditional process is also Markovian. Hence, $p_t(\mathbf{x} |D)$
itself fulfils a Master equation, where the original rates
$f(\mathbf{x}' |\mathbf{x})$ are replaced by time dependent rates
$g_t(\mathbf{x}'|\mathbf{x})$ which take the observations into
account. By differentiating $p_t(\mathbf{x} |D)$ with respect to time
$t$, using the backward Eq.~(\ref{eq:backward}) and the forward
Eq.~(\ref{eq:master}), the rates of the conditional process are found
to satisfy
\begin{equation}
  \frac{g_t(\mathbf{x}' | \mathbf{x})}{f(\mathbf{x}' | \mathbf{x})} =
  \frac{r_t(\mathbf{x}')}{r_t(\mathbf{x})} \,.
  \label{eq:postrate}
\end{equation}
Note that for times $t > t_N$ after the last observation,
$g_t(\mathbf{x}' | \mathbf{x}) = f(\mathbf{x}' | \mathbf{x})$. Hence,
the statistical inference problem requires the repeated solution of a
matrix linear differential equation with dimensionality of the order
$M^d$, where $M$ is the typical number of states accessible to each
species in the system.

We next present an approximate solution to this inference problem
which follows van Kampen's idea of an asymptotic system size expansion
\cite{vanKampen:1981:SPP}. This was developed to approximate the
solution of the Master equation (\ref{eq:master}) in the limit when
the typical number of individuals (molecules) is a macroscopic
quantity and fluctuations are expected to be small. To lowest order, a
classical deterministic rate equation for the process is obtained. The
next order includes fluctuations which obey a Gaussian diffusion
process of the Ornstein-Uhlenbeck type. The method is asymptotically
valid under the conditions of a stable classical equilibrium.
Formally, the two lowest order terms in the expansion can be derived
\cite{Gardiner:1996:HSM} by approximating the jump process first with
a diffusion process using a Kramers-Moyal expansion followed by a
subsequent weak noise limit, where the diffusion term is treated as a
small quantity. In this Letter, we generalize this idea for solving
the backward Eq.~(\ref{eq:backward}) instead. The first step is a
backward Kramers-Moyal expansion \cite{Risken:1984:FPE} in
(\ref{eq:backward}) to obtain a backward Fokker-Planck equation
(BFPE). The corresponding diffusion process has a drift vector
$\mathbf{f}$ and diffusion matrix $\mathbf{D}$ defined by
\begin{eqnarray}
  \mathbf{f}(\mathbf{x}) & = & \sum_{\mathbf{x}' \neq \mathbf{x}}
  f(\mathbf{x}'|\mathbf{x}) (\mathbf{x}' - \mathbf{x}) \,, \\
  \mathbf{D}(\mathbf{x}) & = & \sum_{\mathbf{x}' \neq \mathbf{x}}
  (\mathbf{x}' - \mathbf{x} ) f(\mathbf{x} '|\mathbf{x}) (\mathbf{x}'
  - \mathbf{x})^\top \,.
  \label{eq:defdiffusion}
\end{eqnarray}
We then apply a weak noise expansion by formally rescaling $\mathbf{D}
\rightarrow \epsilon^2 \mathbf{D}$ and assuming that typical state
vectors are close to a nonrandom time dependent state $\mathbf{b}(t)$.
Setting $\mathbf{x} = \mathbf{b}(t) + \epsilon \mathbf{u}$ and
$r_t(\mathbf{x}) = \psi_t(\mathbf{u})$ we expand the BFPE up to order
$\epsilon^2$. The macroscopic state equation
\begin{equation}
  \mathbf{\dot{b}} = \mathbf{f}(\mathbf{b})
\end{equation}
is obtained from terms of order $\epsilon$ whereas from the second
order, we obtain the partial differential equation
\begin{equation}
  \partial_t \psi_t + [\mathbf{A}(t) \mathbf{u}]^\top \nabla \psi_t +
  \frac{1}{2} \mathrm{Tr}\left[ {\mathbf{D}(\mathbf{b}) \nabla
      \nabla^\top} \right] \psi_t = 0
  \label{eq:approx_backward_PDE}
\end{equation}
where $\nabla$ is the vector of derivatives with respect to
$\mathbf{u}$ and $A_{ij}(t) = \left. \partial_{x_j} f_i(\mathbf{x})
\right|_{\mathbf{x} = \mathbf{b}(t)}$. If we assume that the
measurement noise can be modelled by a Gaussian, i.e., $p(\mathbf{y} |
\mathbf{x} ) \propto \exp[- \|\mathbf{y} - \mathbf{x} \|^2 / (2
\sigma^2)]$, a simple solution to (\ref{eq:approx_backward_PDE}) which
is compatible with the jump conditions (\ref{eq:jumpcond}), yields
\begin{equation}
  r_t(\mathbf{x}) \approx \frac{z(t)}{\sqrt{|\mathbf{S}|}} \exp
  \left[-\frac{1}{2} (\mathbf{x} - \mathbf{b})^\top \mathbf{S}^{-1}
    (\mathbf{x} - \mathbf{b})\right] \ ,
  \label{solbackw}
\end{equation}
where the matrix $\mathbf{S}$ satisfies the differential equation
\begin{equation}
  \mathbf{\dot{S}} = \mathbf{A} \mathbf{S} + \mathbf{S}
  \mathbf{A}^\top - \mathbf{D}(\mathbf{b})
\end{equation}
and the normalization ($r$ is not a probability) obeys $\dot{z}(t) =
z(t) \, \mathrm{Tr}[\mathbf{A}(t)]$. Note, that jump conditions lead
to discontinuities for both $\mathbf{b}$ and $\mathbf{S}$ at the times
of the observations. Hence, to compute our approximate solution $r_t$
of the backward equation, we only have to solve the ordinary
differential equations for $\mathbf{b}$ and $\mathbf{S}$ backward in
time starting at the last observation $\mathbf{b}(t_N) = \mathbf{y}_N$
and $\mathbf{S}(t_N) = \sigma^2 \mathbf{I}$, where $\mathbf{I}$ is the
unit matrix. In contrast to the exact backward equation, the
dimensionality of these equations is only of the order $d^2$. In this
approximation, the likelihood $p(D | \theta)$ is given by $p(D |
\theta) \approx (2 \pi)^{d/2} z(0)$ where we have assumed an
uninformative flat distribution for $p_0(\mathbf{x})$.

Strictly speaking, the weak noise expansion for the backward equation
is only valid when the observations $\mathbf{y}_i$ are sufficiently
close to the classical path $\mathbf{b}$. Otherwise, the assumption of
Gaussian fluctuations implicit in our approximation would not be
correct, and large deviation techniques would be required. However,
our likelihood based approach of parameter estimation aims at making
observations highly probable with respect to model parameters. This
will drive typical paths sufficiently close to the observed data to
justify the approximation for likely parameter values.

Although (\ref{eq:approx_backward_PDE}) can be viewed as the backward
equation for a linear stochastic differential equation, our
linearization differs from that used in the well known extended Kalman
smoother \cite{Haykin:2001:KFN} approach of data assimilation which
has a similar computational complexity. This is based on a
linearization applied to the forward filtering process $p_t(\mathbf{x}
| D_{< t})$ which requires the knowledge of initial conditions. Our
backward approach can deal more robustly with vague initial conditions
$p_0(\mathbf{x})$.

The Kalman approach would simply approximate the conditional
probability $p_t(\mathbf{x} | D)$ by that of the linearized model
estimated in the filtering step. In contrast, we apply the combination
of Kramers-Moyal and weak noise expansions a second time to a
(forward) process with jump rate $g_t(\mathbf{x}' | \mathbf{x})$ given
in (\ref{eq:postrate}) leading to a new linearization at a different
state. The drift of the resulting diffusion process is obtained by
expanding $r_t(\mathbf{x}')$ to first order around $\mathbf{x}$ in
Eq.~(\ref{eq:postrate}). Using (\ref{solbackw}) and the definition of
the diffusion matrix (\ref{eq:defdiffusion}), we obtain
\begin{equation}
  \mathbf{g}(\mathbf{x},t) \approx \mathbf{f}(\mathbf{x}) -
  \mathbf{D}[\mathbf{b}(t)] \mathbf{S}^{-1}(t) [\mathbf{x} -
  \mathbf{b}(t)] \,.
  \label{eq:newdrift}
\end{equation}
The subsequent weak noise expansion leads to a Gaussian approximation
for $p_t(\mathbf{x}|D)$ where the mean state vector $\mathbf{m}$ and
and the covariance matrix $\mathbf{C}$ evolve according to
\begin{equation}
  \mathbf{\dot{m}} = \mathbf{g}(\mathbf{m}) \qquad \mathbf{\dot{C}} =
  \mathbf{H} \mathbf{C} + \mathbf{C} \mathbf{H}^\top +
  \mathbf{D}(\mathbf{m})
\end{equation}
with $H_{ij}(t) = \left. \partial_{x_j} g_i(\mathbf{x})
\right|_{\mathbf{x} = \mathbf{m}(t)}$. The discontinuities in the
drift $\mathbf{g}(\mathbf{x}, t)$ at the observation times lead to
discontinuities in the first derivative of $\mathbf{m}$ and
$\mathbf{C}$. When no explicit prior knowledge of the initial state is
known, we start the forward equation with the most likely value
$\mathbf{m}(0) = \mathbf{b}(0)$ and with the uncertainty given by
$\mathbf{C}(0) = \mathbf{S}(0)$.

\begin{table}
  \centering
  \caption{Maximum likelihood estimates of reaction constants based on
    $11$ observations with $\sigma = 1$. Mean values and standard
    deviations have been obtained by averaging over $1000$ samples
    from the Lotka-Volterra process.}
  \vspace{1ex}
  \begin{tabular}{cccc}
    \hline \hline
    parameter & inference result & uncertainty & true value \\
    \hline
    $\alpha$ & $(4.0 \pm 1.2) \times 10^{-3}$ &
    $(0.9 \pm 0.3) \times 10^{-3}$ & $4 \times 10^{-3}$ \\
    $\beta$  & $(2.1 \pm 0.9) \times 10^{-4}$ &
    $(0.5 \pm 0.4) \times 10^{-4}$ & $2 \times 10^{-4}$ \\
    $\gamma$ & $(4.5 \pm 4.7) \times 10^{-3}$ &
    $(1.2 \pm 0.7) \times 10^{-3}$ & $4 \times 10^{-3}$ \\
    $\delta$ & $(1.9 \pm 0.6) \times 10^{-4}$ &
    $(0.5 \pm 0.2) \times 10^{-4}$ & $2 \times 10^{-4}$ \\
    \hline \hline
  \end{tabular}
  \label{tab:lvp}
\end{table}

We have successfully applied our method to observations generated by
simulating two rather simple, but non-trivial reaction systems. The
first one, the well-known Lotka-Volterra model \cite{Boys:2008:BID,
  Gilioli:2008:BIF}, consists of two interacting species,
traditionally named preys and predators. In this system four reactions
are possible, as both preys $X_1$ and predators $X_2$ can be created
or destroyed at any point in time. The reactions and rate laws are
given by
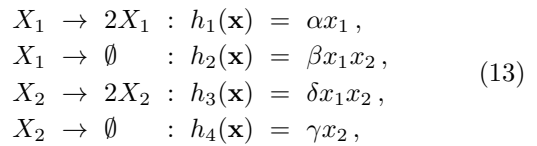
\begin{equation}
  \begin{array}{rclcrcl}
    X_1 &\rightarrow& 2 X_1     &:& h_1(\mathbf{x}) &=& \alpha x_1 \,,
    \\
    X_1 &\rightarrow& \emptyset &:& h_2(\mathbf{x}) &=& \beta x_1 x_2
    \,, \\
    X_2 &\rightarrow& 2 X_2     &:& h_3(\mathbf{x}) &=& \delta x_1 x_2
    \,, \\
    X_2 &\rightarrow& \emptyset &:& h_4(\mathbf{x}) &=& \gamma x_2 \,,
  \end{array}
  \label{eq:lvp}
\end{equation}
where $x_1$ denotes the number of preys and $x_2$ the number of
predators. The second reaction system is a simple genetic
autoregulatory network \cite{Opper:2007:VIM}, which can be found as
parts of the transcriptional regulatory network in biological cells.
It again consists of two species, mRNA $X_1$ and protein $X_2$, and
four reactions. Their associated rate laws are given by
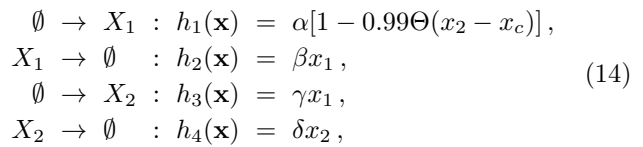
\begin{equation}
  \begin{array}{rclcrcl}
    \emptyset &\rightarrow& X_1 &:& h_1(\mathbf{x}) &=& \alpha [1 -
    0.99 \Theta(x_2 - x_c)] \,, \\
    X_1 &\rightarrow& \emptyset &:& h_2(\mathbf{x}) &=& \beta x_1 \,,
    \\
    \emptyset &\rightarrow& X_2 &:& h_3(\mathbf{x}) &=& \gamma x_1 \,,
    \\
    X_2 &\rightarrow& \emptyset &:& h_4(\mathbf{x}) &=& \delta x_2 \,,
  \end{array}
  \label{eq:gan}
\end{equation}
where $\Theta(x)$ is the Heaviside step function. Both mRNA and
proteins decay exponentially, and proteins are produced by translation
of mRNA with a rate proportional to the mRNA number $x_1$. In order to
regulate the concentration of the protein, the production of mRNA is
down-regulated by a factor $0.01$ as soon as $x_2$ increases beyond a
critical threshold $x_c$.

Both reaction systems have been simulated using Gillespie's algorithm
\cite{Gillespie:1976:GMN}. Observations have been obtained at regular
intervals and corrupted by Gaussian noise with standard deviation
$\sigma$. The exact initial conditions, $x_1(0)$ and $x_2(0)$, have
not been used for inference purposes.

\begin{figure}
  \includegraphics[width=0.45\textwidth]{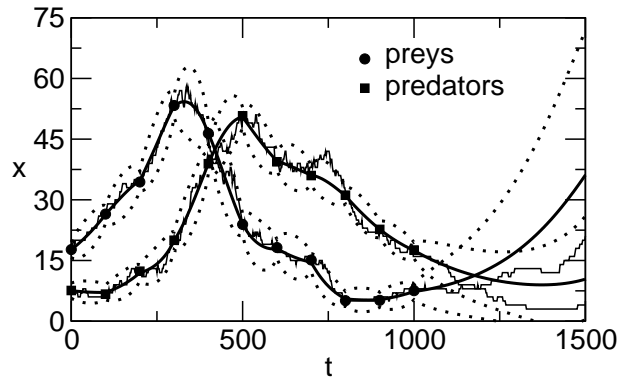}
  \caption{Inference results for a Lotka-Volterra process with
    parameters given in Table \ref{tab:lvp}. Symbols denote noisy
    observations, thick lines show the posterior mean, and the $95\%$
    confidence region for parameters fixed at their maximum-likelihood
    values is surrounded by dotted lines. For comparison, the original
    process is plotted as thin line. Parameter estimates for this data
    set: $\alpha = (6.1 \pm 1.3) \times 10^{-3}$, $\beta = (2.6 \pm
    0.5) \times 10^{-4}$, $\gamma = (4.1 \pm 0.9) \times 10^{-3}$,
    $\delta = (1.8 \pm 0.4) \times 10^{-4}$.}
  \label{fig:lvp}
\end{figure}

As shown in Table \ref{tab:lvp} and Table \ref{tab:gan}, the inferred
reaction constants are reasonable close to their true values.
Additionally, our method is quite fast, because only one backward
integration is needed in each step of the maximization algorithm. In
fact, its computational complexity is nearly independent of the number
of observations.

In the case of the Lotka-Volterra model, we used a standard
gradient-based method for optimization. From the curvature of the
log-likelihood, we also get predictions for the uncertainty of the
parameter estimates (shown in the third column in table \ref{tab:lvp})
which compare fairly well to the average estimation errors obtained
from the simulations.

The success of our method for the case of the genetic autoregulatory
network is somewhat more surprising. The discontinuity of the
reactions as a function of the threshold $x_c$ makes this a highly
nonlinear model. For this case, we have used the Nelder-Mead simplex
method \cite{Nelder:1965:SMF} for parameter optimization which works
without computing derivatives.

\begin{figure}
  \includegraphics[width=0.45\textwidth]{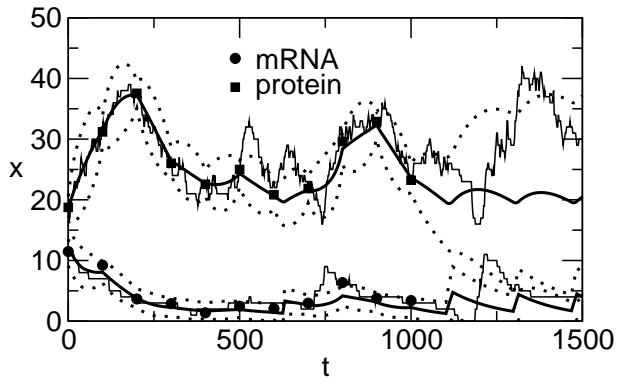}
  \caption{Inference results for the genetic autoregulatory network
    with parameters given in Table \ref{tab:gan}. Symbols and lines as
    in Fig.~\ref{fig:lvp}. Results of parameter estimation for this
    data set: $\alpha = 1.9 \times 10^{-1}$, $\beta = 6.8 \times
    10^{-3}$, $\gamma = 3.0 \times 10^{-2}$, $\delta = 4.2 \times
    10^{-3}$, $x_c = 19.7$.}
  \label{fig:gan}
\end{figure}

Figure \ref{fig:lvp} shows results of state inference for a single
Lotka-Volterra process which is based on the maximum likelihood
parameter estimates. Similar results for the genetic autoregulatory
network are shown in Fig.~\ref{fig:gan}. Here, it is even possible to
predict the strongly nonlinear behavior of the system for time points
after the last observation qualitatively. Further analysis of the
approximate posterior distribution $p_t(\mathbf{x}|D)$ indicates that
the prediction is indeed well calibrated; i.e., $\mathbf{m}(t)$ and
$\sigma_j(t) = \sqrt{C_{jj}(t)}$ are good estimators for
$\mathbf{x}(t)$ and its fluctuations. Consequently, our method gives
reliable information about the state of a reaction system, although
the internal noise is rather large due to the small number of
individuals (molecules) in the examples given here.

\begin{table}
  \centering
  \caption{Maximum likelihood estimates of reaction constants
    based on $11$ observations with $\sigma=1$. Mean values and
    standard deviations have been obtained by averaging over $1000$
    samples from the genetic autoregulatory network model.}
  \vspace{1ex}
  \begin{tabular}{ccc}
    \hline \hline
    parameter & inference result & true value \\
    \hline
    $\alpha$ & $(1.8 \pm 1.2) \times 10^{-1}$ & $2 \times 10^{-1}$ \\
    $\beta$  & $(5.7 \pm 1.9) \times 10^{-3}$ & $6 \times 10^{-3}$ \\
    $\gamma$ & $(4.6 \pm 1.2) \times 10^{-2}$ & $5 \times 10^{-2}$ \\
    $\delta$ & $(7.4 \pm 2.3) \times 10^{-3}$ & $7 \times 10^{-3}$ \\
    $x_c$    & $17.6 \pm 3.6$ & $20$ \\
    \hline \hline
  \end{tabular}
  \label{tab:gan}
\end{table}

Up to now, we have only considered complete observations, which
provide a measurement for each component of $\mathbf{x}$. But often it
is difficult or even impossible to observe all species of a reaction
system. In this case, the last partial observation $\mathbf{y}_N$ is
not sufficient for initializing $\mathbf{b}$ and $\mathbf{S}$.
Therefore, we define an additional virtual observation
$\mathbf{\hat{y}}$ at $t_N$ which contains only noise-free data for
the hidden species. Then, the backward integration can be started with
$r_{t_N}(\mathbf{x}) = p(\mathbf{y}_N|\mathbf{x})
p(\mathbf{\hat{y}}|\mathbf{x})$ and the total likelihood is given by
$p(D | \theta, \mathbf{\hat{y}}) = p(D, \mathbf{\hat{y}} | \theta) /
p(\mathbf{\hat{y}} | \theta)$. Here, we choose $p_{t_N}(\mathbf{x})$,
the marginal distribution of the prior process, as prior distribution
for our virtual observation $\mathbf{\hat{y}}$ so that it and the
denominator of the likelihood cancel out exactly. Therefore, it is
possible to estimate the parameters and to reconstruct the dynamics of
unobserved species by just maximizing $p(D, \mathbf{\hat{y}} |
\theta)$ with respect to $\mathbf{\hat{y}}$ and $\theta$.

We are planning to combine our method with a Monte Carlo approach in
order to sample from the exact posterior. Using our approximation, we
expect to generate good proposal paths for an efficient Metropolis
sampler.

\bibliography{paper}

\end{document}